\documentclass[conference]{IEEEtran}
\IEEEoverridecommandlockouts

\ifCLASSINFOpdf
\usepackage[pdftex]{graphicx}
	\graphicspath{{Figures/}}
	\DeclareGraphicsExtensions{.jpg}
\else
\fi
\usepackage[cmex10]{amsmath}
\usepackage{amsfonts}
\usepackage{xcolor,colortbl}

\usepackage{stfloats}

\begin{document}
\title{Towards Very Large Aperture Massive MIMO: \\a measurement based study}
\author{\IEEEauthorblockN{{\`A}lex Oliveras Mart{\'i}nez,
Elisabeth De Carvalho,
Jesper {\O}dum Nielsen}
\IEEEauthorblockA{Faculty of Engineering and Science, Dept. of Electronic Systems, APNet section\\
Aalborg University,
Aalborg, Denmark\\ Email: \{aom,edc,jni\}@es.aau.dk}}

\maketitle

\begin{abstract}
Massive MIMO is a new technique for wireless communications that claims to offer very high system throughput and energy efficiency in multi-user scenarios. The cost is to add a very large number of antennas at the base station. Theoretical research has probed these benefits, but very few measurements have showed the potential of Massive MIMO in practice. We investigate the properties of measured Massive MIMO channels in a large indoor venue. We describe a measurement campaign using 3 arrays having different shape and aperture, with 64 antennas and 8 users with 2 antennas each. We focus on the impact of the array aperture which is the main limiting factor in the degrees of freedom available in the multiple antenna channel. 
We find that performance is improved as the aperture increases, with an impact mostly visible in crowded scenarios where the users are closely spaced. We also test MIMO capability within a same user device with user proximity effect. We see a good channel resolvability with confirmation of the strong effect of the user hand grip. At last, we highlight that  propagation conditions where  line-of-sight is dominant can be favourable. 
\end{abstract}

\IEEEpeerreviewmaketitle

\section{Introduction}
In the seminal work of Marzetta \cite{Marzetta}, a massive MIMO (Multiple-Input Multiple-Output) system refers to a multi-user MIMO communication system where a base station comprises a very large number of antennas, much larger than the number of served users. In this under-determined multi-user system, the extra spatial Degrees of Freedom (DoF) are exploited to make the multi-antenna multi-user MIMO channel asymptotically orthogonal. In addition, relying on the knowledge of the channel, a proper processing at the Base Station (BS) averages out the fading at the receivers in the downlink direction and at the BS in the uplink direction. Based on those features, enabled by the extra DoF, massive MIMO is recognized as a promising technology for very high system throughput and energy efficiency.

In massive MIMO systems, the spatial DoFs available in the multi-antenna multi-user channel play a central role. With an independent and identically distributed (i.i.d.) modelling of fading, the number of DoF is simply defined and limited by the number of antennas. This simple modelling provides an inappropriate account of the DoF limitation though. For a given propagation environment, the array physical characteristics, its physical size and geometry, define the number of DoF as it defines the angular resolvability. This can be easily understood in a line-of-sight environment, where users can be separated if they are further apart than the resolution unit. In a scattering environment, the array dimension defines the degree of resolvability of the scattering clusters and hence the DoF (see e.g. \cite{Tse}). In this paper, {\it array aperture} refers to the dimension of the arrays. We have two types of array, a linear array and a square array:
the  array aperture is the length of a linear array and the length of the side of a square array.

Increasing the number of antennas within a fixed array aperture is useful to grab all the available DoF up to the limit imposed by the aperture. After this limit is exceeded, increasing the number of antennas does not bring improvement. There is another important point that advocates for very large aperture arrays: as pointed out in \cite{MeasProp} \cite{MassiveMIMO} an increased aperture implies an extended vision range of the environment. The more the array can capture of his environment, the more diversity it can capture, implying more DoF.

The importance of the aperture to achieve the theoretical performance of massive MIMO systems is the motivation for a series of channel measurements performed at Aalborg University where the primary focus is on the impact of the array aperture. The measurements presented in this paper targets a deployment of massive arrays that differs from the current mainstream of a cellular deployment where the massive BS is placed outdoors. In general, we address a deployment in large venues, possibly indoors, where massive arrays with a Very Large Aperture (VLA)  are designed as an integral part of a new large infrastructure.
Referring to the 5G scenarios defined within the EU project METIS \cite{METIS}, VLA arrays can be deployed along walls or ceiling of a shopping mall or airport, around the structure of a football stadium or along the facade of a building. Such scenarios benefit from line-of-sight propagation with the potential of very high rank point-to-point MIMO channels \cite{LoS} and acute discrimination between users, especially in crowd scenarios. They also benefit from rich scattering as the arrays see a wide range of diverse scatterers. Our vision is that the theoretical benefits described in \cite{Marzetta} can be achieved in those large infrastructure deployment, while cellular deployment with BS located in high towers would benefit from the capability of massive MIMO for sharp beams. Our measurement campaign provides a first positive echo to this vision.

To this date, still very few measurement campaigns with published results exist. There are 5 of them \cite{MeasProp} \cite{MassiveMIMO} \cite{LinearPreMeasMIMO} \cite{ChannelMeas} \cite{ChanMeasat}. All target outdoor scenarios at frequency 2.6~GHz. Following the first publications on massive MIMO, the major stress of those measurements is set on the impact of number of antennas, not relating it directly to array aperture. Only in \cite{MeasProp} \cite{MassiveMIMO} can we find a  comparison between two different aperture arrays where the larger aperture array was found to have better performance. 
In \cite{LinearPreMeasMIMO} \cite{ChannelMeas}, the orthogonality between the channels of 2 users  (the normalized scalar product)  has been measured to increase with the number of BS antennas. All the existing measurement campaigns confirm the promises held by the theoretical studies. However, as pointed out in \cite{LinearPreMeasMIMO} \cite{ChannelMeas}, a saturation is observed creating a gap with the performance of i.i.d. channels, likely coming from the limitation of the DoF.
All those campaigns face the problems induced by the large number of antennas and the impossibility to achieve simultaneous measurements with the current technology. The solution adopted is to create a virtual massive array and/or virtual multi-user set-up. Measurements are performed with a set of small number of antennas and users, which are moved for a subsequent measurement. This methodology creates heavy constraints of the measurement protocols imposing the environment to remain static within a whole measurement interval. 

\begin{figure}[!h]
\centering
\includegraphics[width=3.5in]{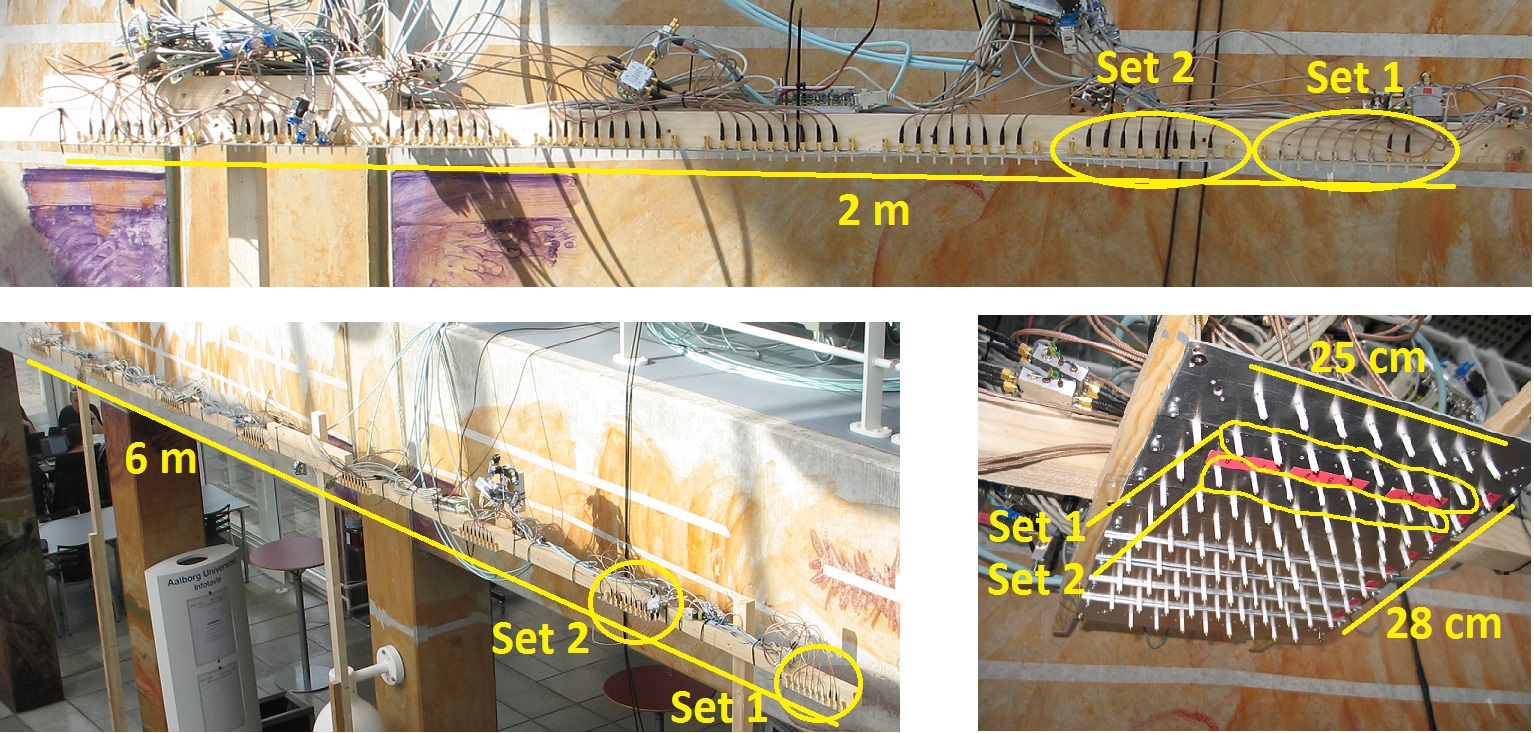}
\caption{Three array shapes are tested with different aperture: Very Large Aperture at the bottom-left, Large Aperture at the top and Compact 2D at the bottom right.}
\label{fig_AllArrays}
\end{figure}

Our measurement campaign involves 64 antennas that are rearranged in 3
different geometrical forms (see fig.~\ref{fig_AllArrays}). In the
first array, the antennas, separated by half the wavelength, are
placed within a square. We refer to this array as “Compact 2D”. It
corresponds to a common view on the design of massive arrays, where
the massive array is as compact as possible. In the “Large Aperture
array”, the 64 antennas are spread along a line of 2 meters. In the
``Very Large Aperture array'', the antennas are spread along a line of 6
meters. Note that one advantage of the compact 2D array is to offer
beamforming capabilities in 2 dimensions: this capability will be
tested against the one-dimensional beamforming (in LoS) offered by the
linear arrays.  The distinctive features of our measurements can be
described as follows: First, we have a multi-user set-up where 8 users
transmit simultaneously to the massive BS.  Second, we test MIMO
capabilities within the same device with and without user proximity
effects: the 8 users hold devices with 2 antennas. Last, we perform
quasi-simultaneous measurements: the set-up includes 16 fully parallel
transmitters (8 mobile units with 2 antennas) and 8 fully parallel
receive units. At the BS, the received signals from 8 antenna elements are
measured simultaneously. Using fast switching, the parallel system is
extended to allow measurements with 64 element BS arrays.

Our general conclusion is that performance is better as the array
aperture increases. More specifically, our main findings are
summarized as follows:
\begin{itemize} 

\item Inter-user link orthogonality: the impact of the aperture is mainly visible when the users are closely grouped. As the number of users increases, the very large array is able to hold performance that is  closed to the i.i.d.  channel. 

\item Intra-user link orthogonality (MIMO capability): we see a good channel resolvability that is better for  the very large array. Furthermore, our measurements  confirm the strong effect of the user hand grip.

\item Power variations across the array: the largest variations are seen for the Very Large Aperture array. However, even for the Compact 2D array, we can see power variations larger than 10dB. \end{itemize}

\begin{figure}[!h]
\centering
\includegraphics[width=1.5in]{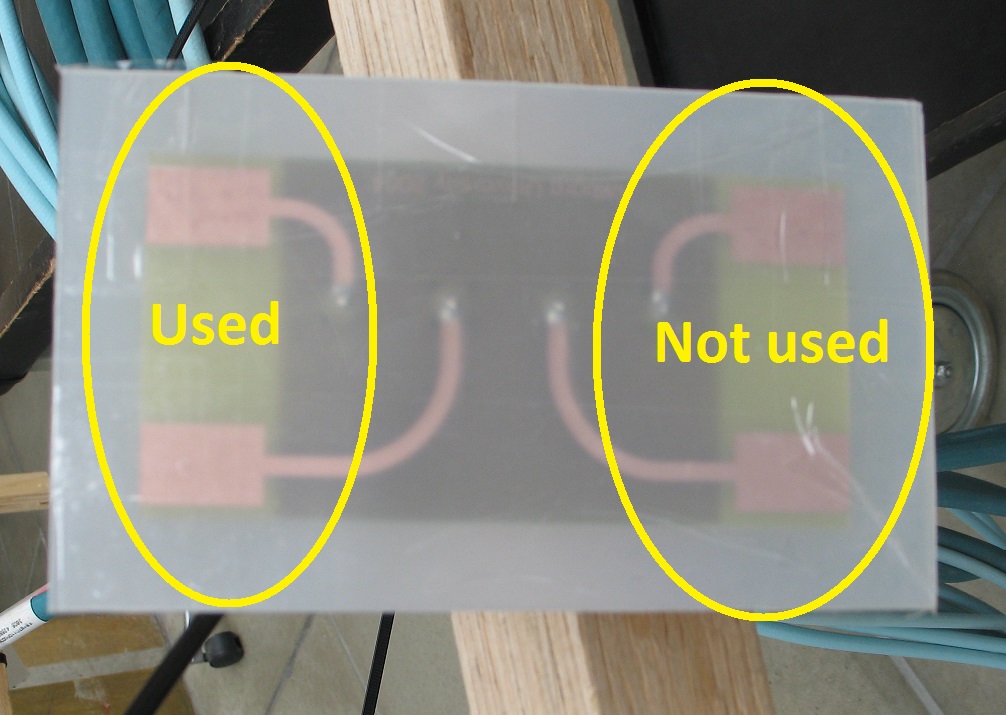}
\caption{Handset with four antennas. Only two are used}
\label{fig_Handset}
\end{figure}

\section{Measurement setup}

\begin{figure}[!t]
\centering
\includegraphics[width=3.5in]{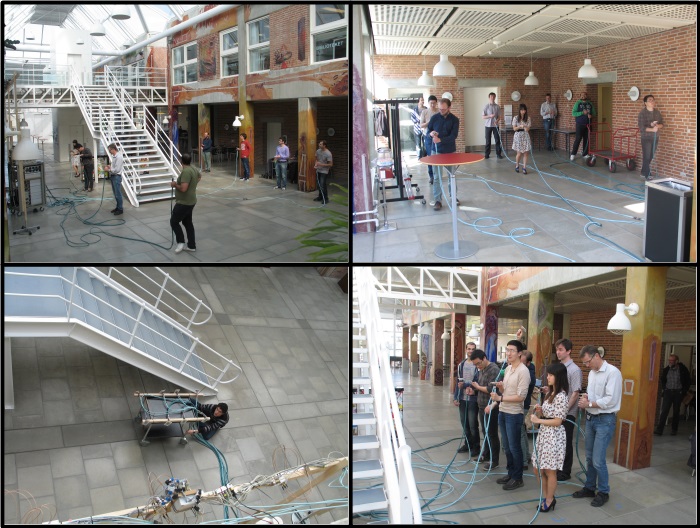}
\caption{Top-left, Spread users LoS. Top-right, spread users NLoS. Bottom-left, Free space in front the stairs. Bottom-right, Grouped users LoS.}
\label{fig_Scenarios}
\end{figure}
The measurements campaign was performed in one of the canteens at  Aalborg University. This location can be considered as a large indoor venue and  has a similar structure as a shopping mall: a big open space with high ceiling, stairs (to go to an upper level), and some small areas on the side with a lower ceiling (see fig. \ref{fig_Scenarios}). 

\subsection{Three array shapes with different aperture}

We use a total of 64 monopole elements. For practical reasons, the 64 elements are grouped in sets of 8 elements. The distance between elements in the same set is $\lambda/2$ at 5.8 GHz. Each set has 2 dummy elements at the ends to provide balanced properties among the active elements (load, coupling, correlation). 
Those sets are arranged in 3 different array shapes, pictured in fig.~\ref{fig_AllArrays}: 1) a square Compact 2D array of 25cmx28cm, where the sets are placed in a square, 2) a Large Aperture linear array of 2 meter long where the sets are placed near to each other along a line, 3) a Very Large Aperture linear array of 6 meter long, where the sets are placed further away. 
The array is placed along a wall of the room, parallel to the stairs.

\subsection{Multi-user Scenarios}
\label{sec_users}

A total of 8 handsets transmit to the massive array (Fig.~\ref{fig_Handset}).
Eight users hold the handsets as if in data mode in one or both hands. The device is located at a few centimeters from the body:  see figure~\ref{fig_Scenarios}. 
The users move randomly in an area of 1 square meter. Our original intent in having this small mobility area was  to measure small scale fading. However, after analyzing the data, we found that power fluctuations originating from the user movement were larger than expected for the measurements to reflect small scale fading. 
As a result, we decided to normalize the channel as described in (\ref{eq:norm}), where the channel at  each measurement is normalized to the same value. The devices are also tested in free space (no user proximity effect) and also moved within a 1 square meter area. 

\begin{figure}[!t]
\centering
\includegraphics[width=3.5in]{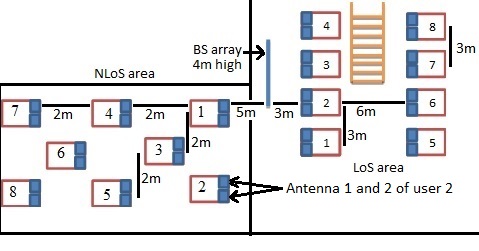}
\caption{Floor map with location of BS array, LoS area, NLoS area and user numbering.}
\label{fig_Users}
\end{figure}

Six scenarios are tested, each one with specific propagation properties, with LoS (Line of Sight) and  NLoS (without LoS) and 
with a specific distribution of the users: see fig.~\ref{fig_Scenarios} and fig.~\ref{fig_Users}. 

\begin{itemize} 
\item \textbf{Spread LoS Parallel}:  the users face the array holding the devices so that the two-antenna array is parallel to the BS array. Four users are in a  line parallel to the BS array between the stairs and the BS array and four of them in a line behind the stairs. The distance between users is 3m and the distance from the users to the array varies from 5m to 12m.

\item \textbf{Spread LoS Perpendicular}: the distribution of users is the same but the users do not face the array but rather 
look in a direction parallel to the array. The two-antenna array in the handsets and the BS array are perpendicular. 
In pure LoS conditions, the two-antenna in the array cannot be likely resolved. 
The purpose of this scenario is to test whether this is the case or whether  scattering is rich enough to allow discrimination between the two antennas. 

\item \textbf{Grouped LoS}: the users are behind the stairs, all in a small area of 1.5mx6m, shoulders to shoulders and moving away from each other. The distance to the array varies from 10m to 12m. 

\item \textbf{Free Space in front the stairs}: the 8 handsets are fastened to a table and the table was moved randomly within a small area. The table is placed between the stairs and the BS array.

\item \textbf{Free Space behind the stairs}: the same as before but the table is placed behind the stairs.

\item \textbf{Spread NLoS}: the users are in non LoS conditions, in the lateral room with a distance between users of 2m and a distance to the BS array from 5m to 11m. 

\end{itemize} 

Figure.~\ref{fig_Users} provides an illustrative floor map of the large room where the measurement were conducted. The stairs are in orange, the BS array is in blue. The 8 users (with 2 antennas each) are depicted in LoS and NLoS conditions. 
To ease the analysis in section~\ref{sec_analysis},  we associate a user number to its geographical position. 
In the scenarios Grouped LoS and Free space, the relative position of the users (and their numbers) is the same but the distance between them is reduced.

\subsection{Equipment and measurement conditions}

A correlation based channel sounder is used to make the
measurements. The carrier frequency is 5.8 GHz and the bandwidth about
100 MHz. There are 16 fully parallel transmitters (the 8 devices with
the 2 antennas), and 8 fully parallel receivers.  A fast switching
mechanism between the receive sets (8 time switch) allow for a
capture of the complete $64\times 16$ MIMO channel in 655$\mu$s.  For each
deployment of the BS array and scenario, 1200 realizations of the
channel were recorded in 20s, while the users move inside the 1 square
meter area.  The measurements are calibrated up to the antenna
connectors, i.e.\ the antennas are considered part of the channel.
Therefore, any mutual coupling and non-ideal characteristics of the
arrays are included, as they would be in an actual system.   Our
analysis is based on narrow band channel data obtained via Fourier
transforms of the wideband measurements. The statistics are taken over
the 1200 measurements.

\section{Numerical analysis of the measurements}
\label{sec_analysis}
In this section we focus on the analysis of the results obtained in the measurement campaign. The notations are as follows:  the number of users is $K = 8$, the number of antennas for each user is $N = 2$, the number of base station antennas is $M = 64$, and the number of channel realizations is $R = 1200$. In this paper, we report results for the most significant scenarios. 

The results obtained are compared with the i.i.d. Gaussian channel, to provide a comparison between the real measurements and the theoretical results. At the same time the three different arrays are evaluated and compared to each other to show the effect of the aperture on the channel performance.

To analyze the results, the communication system considered is a single cell MU-MIMO (Multi-user MIMO) system with $K$ users, having $N$ antennas each one, and a base station with $M$ antennas serving the users. We denote $\textbf{H}_{r}\in\mathbb{C}^{M \times KN}$ the matrix corresponding to   realization $r$ of the channel. 
Later some subsets of this matrix will be considered. Notice that $M > (KN)$.
We denote 
$\textbf{h}_{kr}^{(n)} \in\mathbb{C}^{M \times 1}$ as the channel vector from antenna $n \in \{1,2\}$ in the handset of user $k$ to the BS array. In matrix $\textbf{H}_{r}$,  the two user channel vectors are placed in two consecutive columns. 
The system has a power gain control so all the users receive the same power regardless their distance to the base station.  The power of channel vector $\textbf{h}_{kr}^{(n)}$  is normalized: 
\begin{equation}
\bar{\textbf{h}}_{kr}^{(n)}= \frac{\textbf{h}_{kr}^{(n)}}{\left\|\textbf{h}_{kr}^{(n)}\right\|}\sqrt{M}.
\label{eq:norm}
\end{equation}
In this way it is fair to compare users in different location and the results will be determined by the channel conditions and not by the path loss of the user. We denote  $\bar{\textbf{H}}_{r}\in\mathbb{C}^{M \times KN}$ as the channel matrix made out of the normalized vectors in (\ref{eq:norm}).
Several figures of merit are obtained from the channel matrix giving a basis to understand the channel.

\subsection{Overview of the Correlation Properties}

A first step to analyze the behavior of the channel is to look at the correlation between channel vectors, encompassed in the following matrix:
\begin{equation}
\textbf{S}=\frac{1}{R}\sum\limits_{r=1}^R \bar{\textbf{H}}_{r}^H \bar{\textbf{H}}_{r}
\end{equation}

\begin{figure}[!t]
\centering
\includegraphics[width=3.5in]{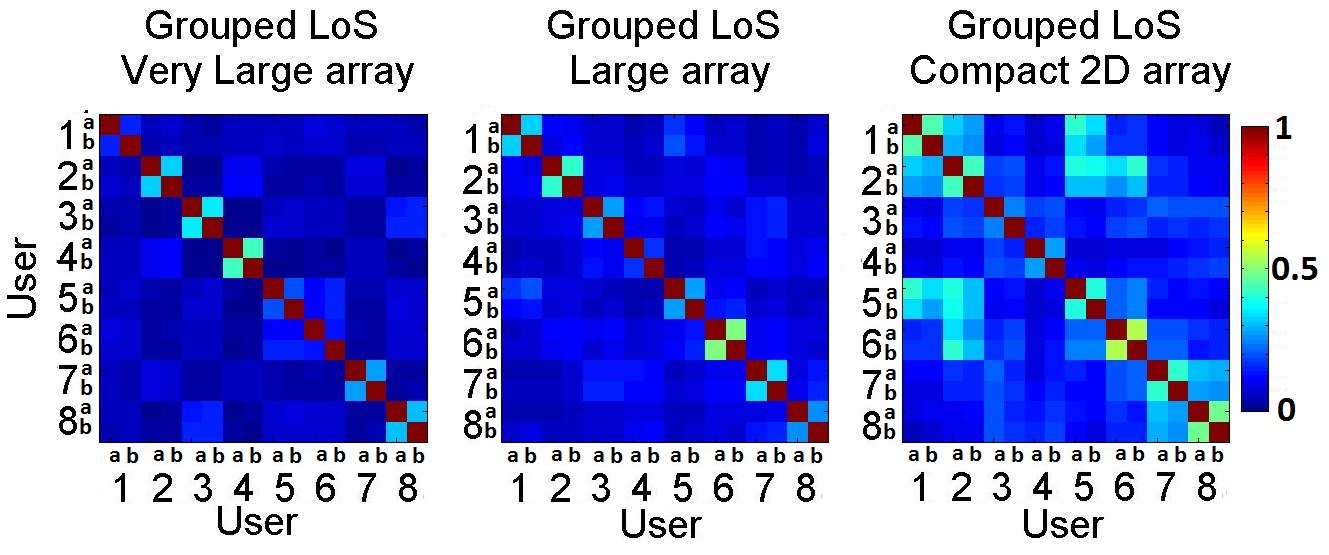}
\caption{Correlation matrix for the scenario: Grouped users in LoS}
\label{fig_SP_GL}
\end{figure}

\begin{figure}[!t]
\centering
\includegraphics[width=3.5in]{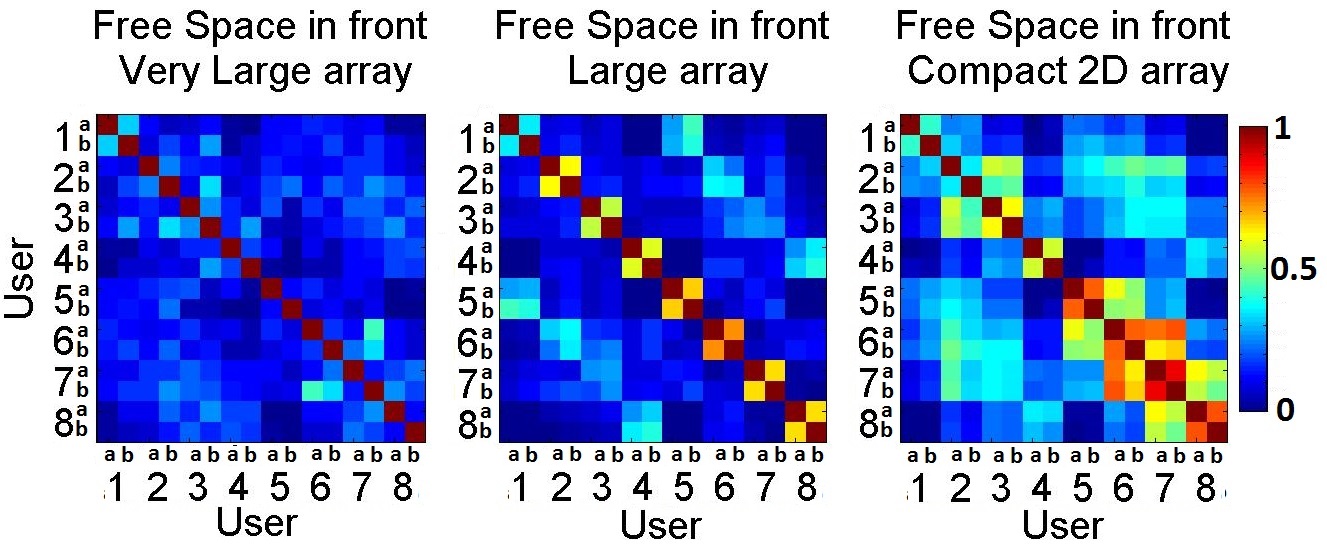}
\caption{Correlation matrix for the scenario: Free space in front of the stairs}
\label{fig_SP_FF}
\end{figure}

The elements of matrix $S$ are pictured in fig.~\ref{fig_SP_GL} and
fig.~\ref{fig_SP_FF}.  The 2x2 blocks along the diagonal represent the
correlation between the channels for the same device. The off-diagonal
blocks represent the correlation between the channels of different
devices. We notice a significant difference in the behavior of the
intra-user and inter-user channels. This is the reason why we later
detail the performance separately for those two kinds of channels.
Matrix $\textbf{S}$ is shown for two scenarios: "Grouped User in LoS"
and "Free space". Both scenarios illustrate a crowd scenario with
devices closer to each other in the "Free space" case. Furthermore, by
comparing both scenarios, user proximity effects can be assessed.  In
the figures, the two antennas associated with each user are denoted
$\textbf{a}$ and $\textbf{b}$, respectively.

{\bf Inter-user properties}: the Very Large Aperture array clearly performs the best and the Compact 2D array clearly performs the worse. While the Very Large Aperture and Large Aperture arrays can satisfactorily discriminate all the users in the grouped LoS scenario, correlation among users appear for the Compact 2D array. The correlations appear not only for users in the same line, but also for users behind each other  (5,1 or 6,2). 
As the devices become more packed (free space), performance degrades for all arrays with the apparition of correlated users. For the Large Aperture array,  strong correlations appear for some pairs of users (1,5 and 2,6 and 3,7 and 4,8) that are behind each other. So we deduce that the Large Aperture array loses its resolution in the elevation angle. 
Finally in the Compact 2D array the correlation is large even for neighboring users (specially the ones further away 
from the base station). 
Channels with low correlation can be seen for users that are far from each other (1,4 and
1,8 and 4,5 and 5,8). 

{\bf Intra-user properties}: It is expected that the channel correlation properties within a same device are worse than across devices as the antenna are close by. In the free space scenario, the Very Large Aperture array performs remarkably better than the other arrays. Comparing now with the "Grouped LoS" scenario, we can notice the large impact of user proximity effects which can  equalize the performance among arrays. 

\subsection{Inter-user properties}

Here the inter-user channel properties are considered. The goal is to explore what happens when the number of users in the system increases. 
We call the number of users in the system $C$ and we increase $C$ from 1 to $K$. We collect all the combinations of $C$ channels from $C$ different users (only one channel per user out of the two channels is selected). We form a new channel matrix for realization $r$ denoted generically as $\textbf{G}_r$ and take statistics on our metric over the different user combinations and channel realizations. 

The metric adopted here is the sum of the eigenvalues normalized by largest eigenvalue (NPCG, Normalized Parallel Channel Gain). 
Denoting $\lambda_{i,r}$ as the $i$th eigenvalue of the matrix $\textbf{G}_{r}^H \textbf{G}_{r}$, then the metric is defined as: 
\begin{equation}
\text{NPCG} =  \frac{1}{\lambda_{\max,r}}\sum\limits_{i=1}^C \lambda_{i,r}.
\label{eq:se}
\end{equation}
$\lambda_{\max,r}$ is the largest eigenvalue. 
This metric is preferred to the condition number of the channel as the latter gives only information on the ratio between the larger and the smaller eigenvalue, whereas metric (\ref{eq:se}) reflects the behavior of all the eigenvalues. Notice that, for a channel that is very well conditioned with equal eigenvalues,  $\text{NPCG} = C$ and for a channel that is poorly conditioned $\text{NPCG} = 1$. 

\begin{figure}[!t]
\centering
\includegraphics[width=3.5in]{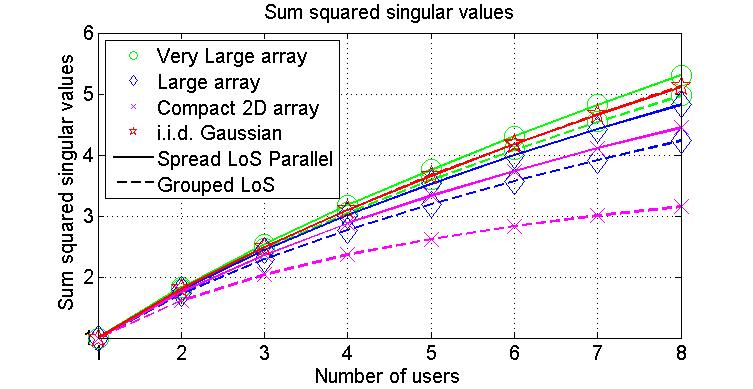}
\caption{Normalized sum of eigenvalues w.r.t. increasing number of users in the Spread LoS with users antennas parallel to the BS array and grouped LoS.}
\label{fig_SE_1}
\end{figure}

In the same figure, fig. \ref{fig_SE_1} two scenarios are represented, 'Spread LoS' with their two antennas parallel to the BS array and 'Grouped LoS'. Comparing both scenarios, it is obvious that the more spread the users, the better. For all the number of users and all the arrays the channel of grouped users has worse conditioning. The limitation of the Compact 2D array in the DoF available to 8 users is particularly visible in the grouped LoS scenario. Note that the Very Large Aperture array in the spread LoS scenario  gives  performance slightly better than the i.i.d. Gaussian case. This can happen in LoS conditions depending on the location of the users. 

In the next figure, fig. \ref{fig_SE_2}, two scenarios are plotted. One is the 'Spread NLoS' and the other is the 'Free space behind the stairs' (the scenario with minimal inter-device distance). The three arrays have a worse channel conditioning than the previous scenario. For the Compact 2D array adding new users can hardly increase the metric showing again a limitation in the DoF. 
Two effects can be observed. On one hand the effect of NLoS propagation and on the other hand the effect of closely spaced devices. In both cases the Very Large Aperture array and the Large Aperture array have a similar behavior. It appears that the scattering is rich enough so that increasing the aperture from large to very large does not bring improvement in the NLoS scenario. 
At last, we show the results for the 
free space scenario, where we can see an overall degradation of the performance. 

\begin{figure}[!t]
\centering
\includegraphics[width=3.5in]{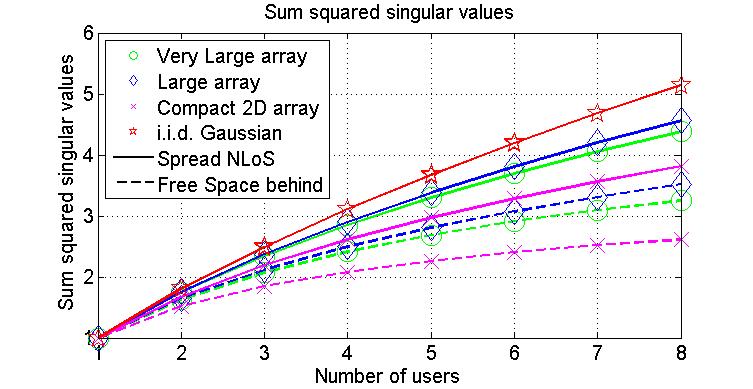}
\caption{Normalized sum of eigenvalues w.r.t. increasing number of users in the Spread NLoS and Free space in front of the stairs scenarios.}
\label{fig_SE_2}
\end{figure}

\subsection{Intra-user properties}

\begin{figure}[!t]
\centering
\includegraphics[width=3.5in]{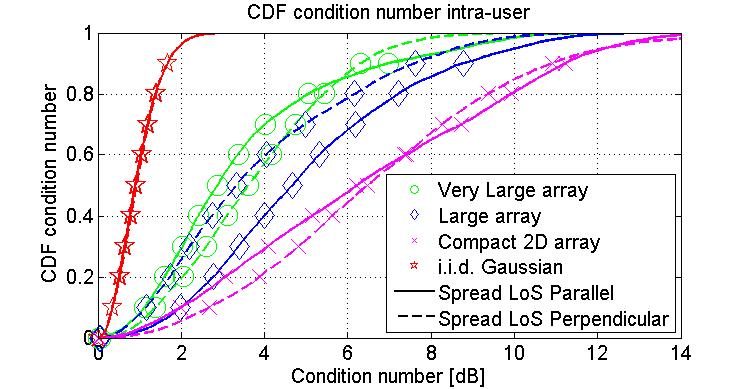}
\caption{Intra-user condition number in the Spread LoS with users antennas parallel and perpendicular to the BS array}
\label{fig_CN_1}
\end{figure}

\begin{figure}[!t]
\centering
\includegraphics[width=3.5in]{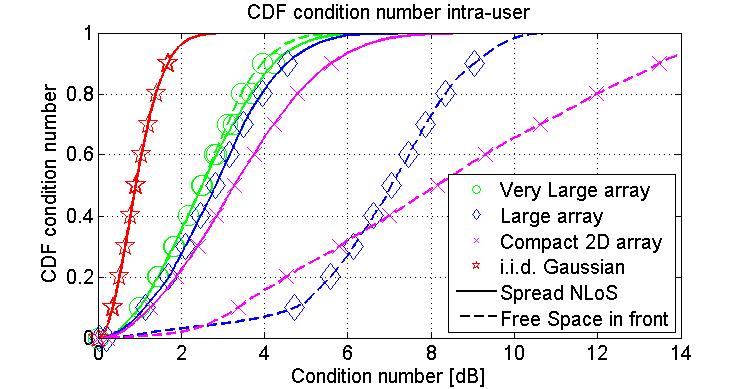}
\caption{Intra-user condition number in the Spread NLoS and Free space in front of the stairs scenarios}
\label{fig_CN_2}
\end{figure}

To evaluate the MIMO channel properties within a same device, we go back to the conventional condition number as we have only 2 antennas. The condition number  indicates how spread the eigenvalues of the channel are. 
It is defined as $\text{CN} = 10\log_{10}(\frac{\lambda_{\max}}{\lambda_{\min}})$ where  $\lambda_{\max}$ and $\lambda_{\min}$ are the maximum and minimum eigenvalues of the MIMO channel matrix. 
Statistics on the condition number are taken over the $R$ realizations of the channel. 

Fig. \ref{fig_CN_1} and Fig. \ref{fig_CN_2} shows the cumulative distribution function (CDF) of the condition number. First, it can be noticed that there is a significant gap between the i.i.d. case and the measured cases, mostly due to user proximity effect and small spacing between the antennas. In general, the Very Large Aperture array still performs the best with slopes that are steeper than the other arrays. As the aperture increases, the number of DoF increases and the distribution of the eigenvalues tend to a deterministic quantity. This is in line with a similar effect in i.i.d. channels when the number of antennas (and hence DoF) increases. 

Fig. \ref{fig_CN_1} accounts for the 'Spread User' scenarios, where the array formed by the antennas at the device is parallel or perpendicular to the BS array. We do not observe a clear tendency in the comparison between both scenarios, while the scenario with perpendicular arrays performs much worse in pure LoS than the scenario with parallel arrays. 
This might  indicate that scattering is rich enough to enable MIMO capabilities in both cases. This observation is actually positive: it makes the access robust to device orientation as  the devices will have a random orientation relative to the BS in general. 

Looking at the fig. \ref{fig_CN_2}, where the users are 'Spread NLoS' and in 'Free space in front of the stairs', we observe that the Very Large Aperture array remains robust towards both scenario conditions.
Both large and compact arrays also give robust performance in the NLoS scenarios brought by a rich scattering environment. 
Poor performance is observed in the LoS free space scenario. More analysis is needed to fully understand this case. Our interpretation is that performance tend to depend on location in LoS, suggesting a more favorable location relative to the Very Large aperture array compared to the other arrays.

\subsection{Power Variations across the Array}

As mentioned in \cite{MeasProp} \cite{MassiveMIMO} \cite{ChannelMeas} where measurements involving a large aperture massive array have been performed, channel characteristics become non-stationary across the array, such as the received power or the direction of arrivals. In general, the environment that is observed is different from one part of the array to another. 

\begin{figure}[!t]
\centering
\includegraphics[width=3.5in]{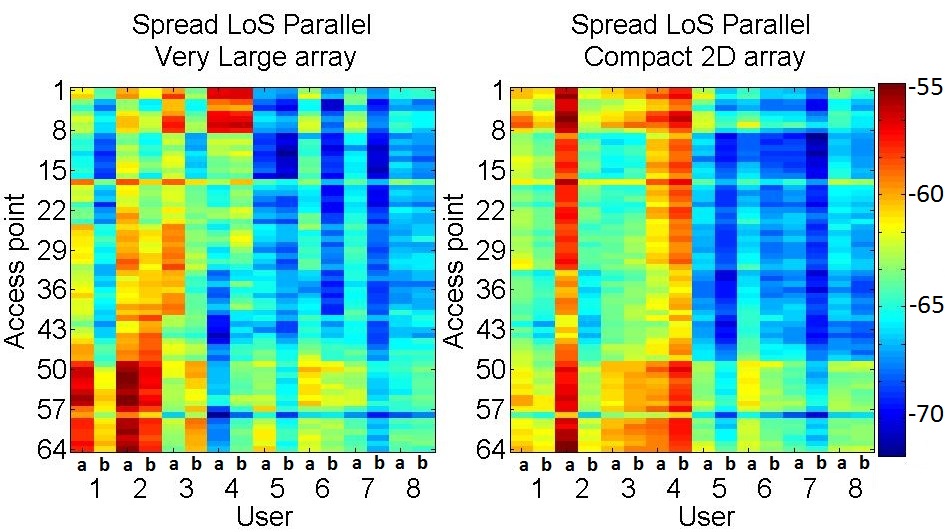}
\caption{Average power variations across the Very Large and the Compact 2D array for the "Spread user in LoS" scenario.}
\label{fig_power}
\end{figure}

In this section, to illustrate this phenomenon, we examine the variations of the received power across the array for one scenario (i.e. 'Spread LoS') and two arrays, the very large and the compact one, shown in fig.~\ref{fig_power}. 
The x axis represents each user (2 antennas per user), and the y axis represents the received power averaged over the 1200 measurements at each of the 64 BS antennas. 
Several observations can be made. 
First, obviously, the users that are further away from the array have a smaller received power. 
Second, as the aperture increases, so does the impact of path loss variations across the array. This is more visible for users that are closer to the array in front of the stairs. Third, the power variations depend on the environment: users 7 and 8 are positioned behind the stairs that have a shadowing effect and make the receive power uniform compared to other users. 
At last, the power variations are smaller in the Compact 2D array case compared with the Very Large Aperture one. However, even for this small aperture, significant variations of 10dB order can be observed. 

The property of non-stationarity across a large aperture array is a new feature that gives a distinctive edge to the type of  communication system studied in this paper. It impacts the performance of the system but also the multi-user access methods which could be beneficial in terms of multi-user diversity. 

\section{Conclusion and Future Work}
The presented investigation describes a measurement campaign involving a massive array with 64 antennas and 8 users with MIMO capabilities in a large 
indoor environment. The main purpose is to investigate on the impact of the massive array aperture. 
Three different shape and aperture of the base station array were tested as well as different propagation conditions (LoS and NLoS), user device distribution (spread and grouped) and user proximity effect. 
The measurements confirm that the aperture is important to create the spatial DoF that brings the benefits promised by the theoretical studies on massive MIMO. 
We found that the array with the largest aperture perform the best with performance close to the i.i.d. channel. The channel tends to be better conditioned bringing very good discrimination among users but also between antennas of a same device, where the user proximity effect still has a major effect. Building on this experience drawn from this first measurement campaign, we are planning a new campaign involving a much  larger number of antennas in a larger venue as well in outdoor conditions. 

\section*{Acknowledgment}

The research presented in this paper was partly supported by the
Danish Council for Independent Research (Det Frie Forskningsr{\aa}d)
DFF–1335–00273. Part of this work has been performed in the framework of
the FP7 project ICT-317669 METIS, which is partly funded by
the European Union. The authors would like to acknowledge
the contributions of their colleagues in METIS, although the
views expressed are those of the authors and do not necessarily
represent the project.

\end{document}